# Modulation of sub-optical cycle photocurrents in an ultrafast near-infrared scanning tunnelling microscope

Andrea Rossetti[1†], Florian Pagnini[1], Majdi Assaid[1], Christoph Schönfeld[2], Alfred Leitenstorfer[2], Markus Ludwig[1,3], Daniele Brida[1*]

[1]Department of Physics and Materials Science, University of Luxembourg, Luxembourg, Luxembourg,
[2]Department of Physics and Center for Applied Photonics, University of Konstanz, Konstanz, Germany
[3]Institute for Advanced Studies, University of Luxembourg, Campus Belval, L-4365 Esch-sur-Alzette, Luxembourg

*daniele.brida@uni.lu

**Abstract:** Lightwave-driven scanning tunnelling microscopy (STM) at near-IR frequencies promises an unprecedented combination of atomic spatial resolution and temporal resolution approaching the attosecond range. To achieve this goal, high-sensitivity optical control and detection of sub-cycle tunnelling currents must be achieved at the STM junction. Here, we demonstrate the generation and detection of coherent ultrafast currents across the junction of an STM illuminated by near-infrared single-cycle pulses. We introduce a novel modulation scheme that avoids time-dependent thermal loading while selectively isolating carrier-envelope phase (CEP)-dependent photocurrents. All artifacts arising from periodic modulation of laser power and thermal coupling are efficiently suppressed, enabling a clean readout of the coherent portion of the ultrafast tunneling current.

## 1. Introduction

The fundamental interactions governing the properties of condensed matter take place at sub-nanometric lengths and characteristic times between the attosecond-scale motion of electronic wavepackets and tens to hundreds of femtosecond lattice dynamics. The synthesis of ultrafast optical transients across wide portions of the electromagnetic spectrum [1] readily allows to investigate a plethora of phenomena, including free-carrier and lattice thermalization in metals[2] [3], excitonic dynamics in semiconductors [4] and spin dynamics in magnetic samples[5]. However, standard all-optical approaches for ultrafast experiments are limited by the spatial resolution set by diffraction. To overcome this, the integration of ultrafast spectroscopy methods with scanning probe techniques has been proposed and partially demonstrated in recent years [6], [7], [8], [9]. In this context, ultrafast scanning tunnelling microscopy (u-STM) and spectroscopy (u-STS) promises to combine atomic spatial with attosecond temporal resolution, providing the ultimate tool to investigate the properties of condensed matter at its natural spatiotemporal coordinates [10].

A first milestone towards u-STS is represented by the lightwave-driven STM[10], whose working mechanism is sketched in figure 1. Single-cycle pulses with stable carrier-envelope phase (CEP) are focused onto the STM junction. The laser electric field transient distorts the potential energy barrier between tip and sample and drives sub-cycle tunnelling currents across the STM junction. While the time integral of the laser electric field is equal to zero, a net time-integrated current flows across the tunnelling contact due its strongly nonlinear response to the applied field. In this picture, the time resolution of the current transient is better than – but still dictated by - the duration of the optical cycle of the pulse, while the magnitude and sign of the net current flow depend on the exact shape of the waveform and can ultimately be controlled via manipulation of the CEP [11]. First experimental implementations of lightwave STM were realized using THz pulses [12]. While the THz-STM is an invaluable tool to investigate lattice dynamics with atomic spatial and sub-picosecond temporal resolution[13], it still lacks the temporal resolution to study faster processes such as wavepacket dynamics within molecules and nanostructures, which typically happen well within 100 fs [14]. To improve the temporal resolution, lightwave-driven STM at higher carrier frequencies is crucial. Recently, an ultrafast STM driven at mid-infrared frequencies attained a time resolution better than 30fs. [15] In this context, the near-infrared spectral region represents a potentially ideal candidate, since it allows to combine optical cycles shorter than 10fs and non-resonant interaction with most solids or molecules. While this possibility has been partially explored [16], evidence for an ultrafast STM driven at near-IR frequencies remains ambiguous. In particular, full spatio-temporal resolution of elementary processes at the atomic scale has not been reached so far due to several drawbacks of present methodology.

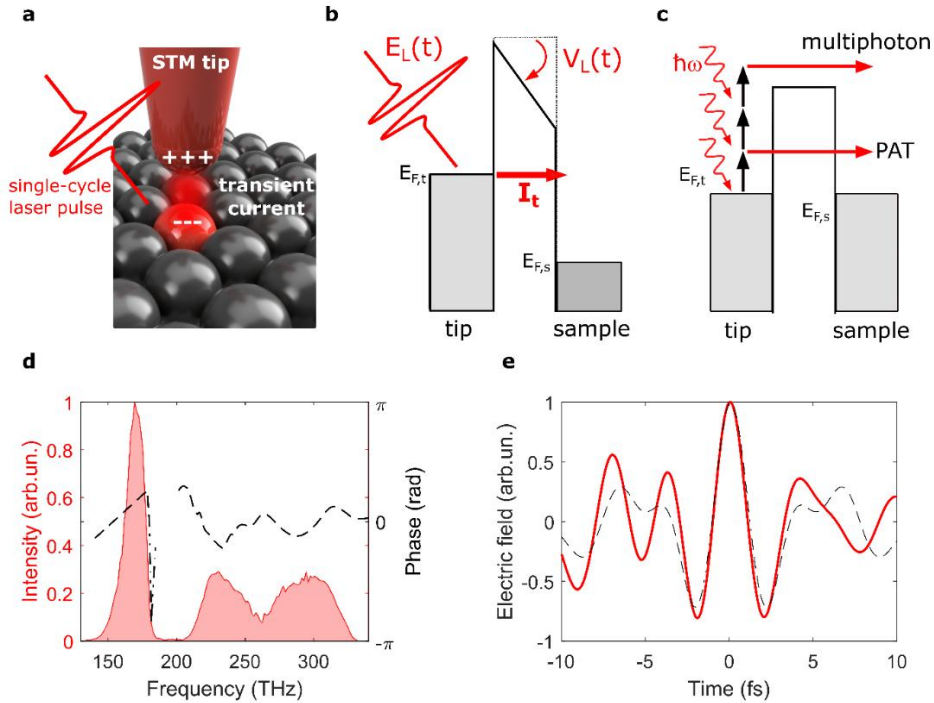

Figure 1: a) Conceptual representation of a lightwave-driven STM. b) Sketch of the potential energy landscape of the process depicted in a). The laser electric field $E_L(t)$ bends the junction barrier by a potential energy difference $eV_L(t)$. The voltage transient causes an imbalance between the Fermi levels $E_{F,t}$ and $E_{F,s}$ of tip and sample, causing a current $I_t$ to flow across the junction within sub-optical-cycle timescales. c) Under illumination at photon energies $\hbar\omega$ in the near-IR spectral range, other processes such as multiphoton photoemission or photo-assisted tunnelling (PAT) contribute to the overall photocurrent. d) Spectral intensity (measured, red shaded area) and phase (obtained via two-dimensional interferometry characterization, black dashed line) of the laser pulses used in this work. e) Temporal profile of the near-IR laser transients (red line) obtained from the spectral intensity and phase of panel (d) for CEP=0. As reference, the dashed black line shows the bandwidth-limited electric field profile at CEP=0.

Synthesis of phase-stable single-cycle pulses centered at 1.5 µm opened the possibility for a lightwave-driven STM at telecom wavelengths [17]. In this wavelength range, sub-femtosecond coherent control of field-driven currents across few-nm plasmonic nanogaps has been demonstrated[12], [13]. Despite this, several conceptual and technical challenges must still be addressed to implement an u-STM at near-IR frequencies. At THz frequencies the effect of the photon energy on the light-junction interaction is negligible. This justifies an adiabatic treatment of ultrafast electron tunnelling driven by THz waveforms, similar to the case of a static bias.

Moreover, high field enhancement factors at THz frequencies allow to attain strong local peak fields above 10 MV/cm at moderate free-space peak fields lower than 1 kV/cm [20], [21]. Furthermore, at frequencies around 200 THz processes like multi-photon photoemission[22], photo-assisted tunnelling (PAT) [23],[24] or hot-electron tunnelling [25] significantly contribute to the overall photocurrent alongside field-driven photoemission from the Fermi level. Overall, the near-IR spectral range features significantly shorter optical cycles and lower field enhancement factors [26], reducing the number of field-driven tunnelling electrons per pulse compared to the THz case. Finally, thermal modulation of the tunneling gap separation caused by near-IR pulses can lead to significant spurious laser-induced currents and deteriorate the stability of the STM junction [27] [28], [29], [30]. All these effects render detection and isolation of coherent STM photocurrents driven by NIR pulses extremely challenging. As recently discussed in a theoretical paper by Borisov et al. [11], coherent laser-driven effects in a tunnelling junction can be identified by CEP-dependent oscillations in the photocurrent, analogously to the case of optical-field-emission from metal surfaces and nanostructures [19], [31], [32]. In this work, we investigate CEP effects in photoemission across an STM junction driven at a carrier frequency >200 THz. Moreover, we demonstrate a way to isolate coherent photocurrents hidden beneath a large phase-independent background, advancing the way towards attosecond scanning tunnelling microscopy.

## 2. Coherent photoemission across the STM junction

The laser system employed in this work is a home-made Er:fiber source operating at a repetition rate of 40 MHz [33], [34]. A schematic representation of the setup is shown in figure 2a. Starting from the amplified output of an Er:fiber oscillator, the CEP of the pulse train is passively stabilized via spectral broadening in a highly-nonlinear germanosilicate fiber (HNF) and subsequent difference frequency generation between the red and blue wings of the spectrum. The CEP-stable output is amplified in a two-stage Er:fiber amplifier (EDFA) and spectrally broadened, again, to an octave-spanning supercontinuum centered at 200 THz in a second HNF. In the following, we will refer to the first amplification stage after CEP stabilization as the "preamplifier". After compression, the laser transients are characterized via two-dimensional shearing interferometry (2DSI) [35]. The spectral intensity and phase are shown in figure 1d. From these parameters, we can retrieve the waveform of the laser electric field transient as shown in figure 1e for CEP = 0. As discussed in previous works [19], [36], the CEP in this laser system can be controlled without affecting the pulse envelope by inserting a wedge pair in the beam path after the DFG crystal. The single-cycle transients are then focused by an off-axis-parabolic mirror (OAPM) located inside the STM chamber onto the junction of a commercial STM (Unisoku). The STM junction consists of a mechanically polished Pt/Ir tip and a mechanically exfoliated highly-oriented pyrolytic graphite (HOPG) substrate sample used as a benchmark. The non-resonant interaction of the incoming near-IR pulses with either the tip or sample material ensures that the temporal waveform of the pulse is maintained at the optical near-field in the gap region. Considering a pulse energy of 500 pJ and a focal length f = 35 mm of the OAPM, we estimate peak free-space electric fields reaching up to approximately 10 MV/cm in the focal plane. Considering a conservative tenfold field enhancement factor at the STM junction[37], we expect that strong-field interaction conditions [38] can be attained at STM gap under the present experimental conditions [11].

As discussed in the introduction, several mechanisms can contribute to photoemission across a STM gap illuminated by single-cycle NIR pulses. For large gap sizes, where beneath-the-barrier contributions are suppressed, we can expect that multiphoton-photoemission and optical-field-emission into vacuum states mainly contribute to the overall photocurrent. Considering the work function of Pt/Ir $\phi \simeq 5.5$ eV [39] and the photon energy at the carrier frequency $h\nu \simeq 0.8$ eV, multi-photon photoemission from the tip would require n = 7 photons. In figure 2b, we show the photocurrent from the isolated tip (large gap size $d_{gap} > 100$ nm) for different pulse energies of the incoming field. We plot the experimental data (filled circles) alongside a power law fit $y = a + bx^n$ with n as free parameter. The best fit (dashed line) yields an exponent of $n \simeq 7$ and suggests that the main photoemission channel from the isolated tip is indeed an above-the-barrier process requiring the absorption of 7 photons.

To investigate the presence of coherent effects, we acquire the photocurrent at fixed pulse energy $E_p = 500$ pJ while sweeping the CEP over more than $3\pi$ while moving one of the wedges. The photocurrent is acquired directly from the transimpedance amplifier (no lock-in detection) and averaged over 30 successive CEP sweeps. To ensure that the beginning of each CEP sweep starts from a well-defined phase value, the initial wedge position is adjusted to restore the f-2f fringe pattern before each scan. The results obtained for the case of photoemission from an isolated tip are shown in the top panel of figure 2c (dark blue circles) and exhibit a small current oscillation with amplitude $I_0 \simeq 35$ fA on top of a positive background of 7.5 pA. From the power-law analysis discussed above, we argue that the large positive offset is due to multi-photon photoemission. More importantly, the period of the current oscillation matches with the CEP shift $\Delta\phi_0$ and demonstrates that a partial coherent optical control of the photocurrent can be achieved. As discussed in [11], CEP-effects at large gap sizes can be attributed to both ponderomotive effects after photoemission and to the onset of optical-field-emission contributions to the overall photocurrent. Photoemission becomes even more complex for small STM gap size $d_{gap} < 10$ nm. As discussed in [23] below-the-barrier PAT [40] and tunnelling from the Fermi level become increasingly relevant at small gap sizes. At the same time, we can expect quiver-motion effects to be quenched at small gaps due the almost instantaneous interaction between the tunnelling electron and the driving field [41].

In this work, we focus on isolating the coherent, carrier-envelope phase (CEP)-dependent portion of the tunneling current, as it may uniquely provide access to tunneling currents at sub-cycle optical time scales. As a first step, we measure the total STM current only driven by the laser at a fixed tunneling gap in the absence of a static bias as a function of CEP. Subsequently, we experimentally investigate how the CEP-dependent contribution to the photocurrent varies with the gap size $d_{gap}$. We start our experiment with the STM in static tunnelling conditions defined by a setpoint current $I_s = 1$ nA and bias $V_0 = 1$ V. These parameters determine an unknown, yet reproducible and well-defined tip-sample distance $z_0$.

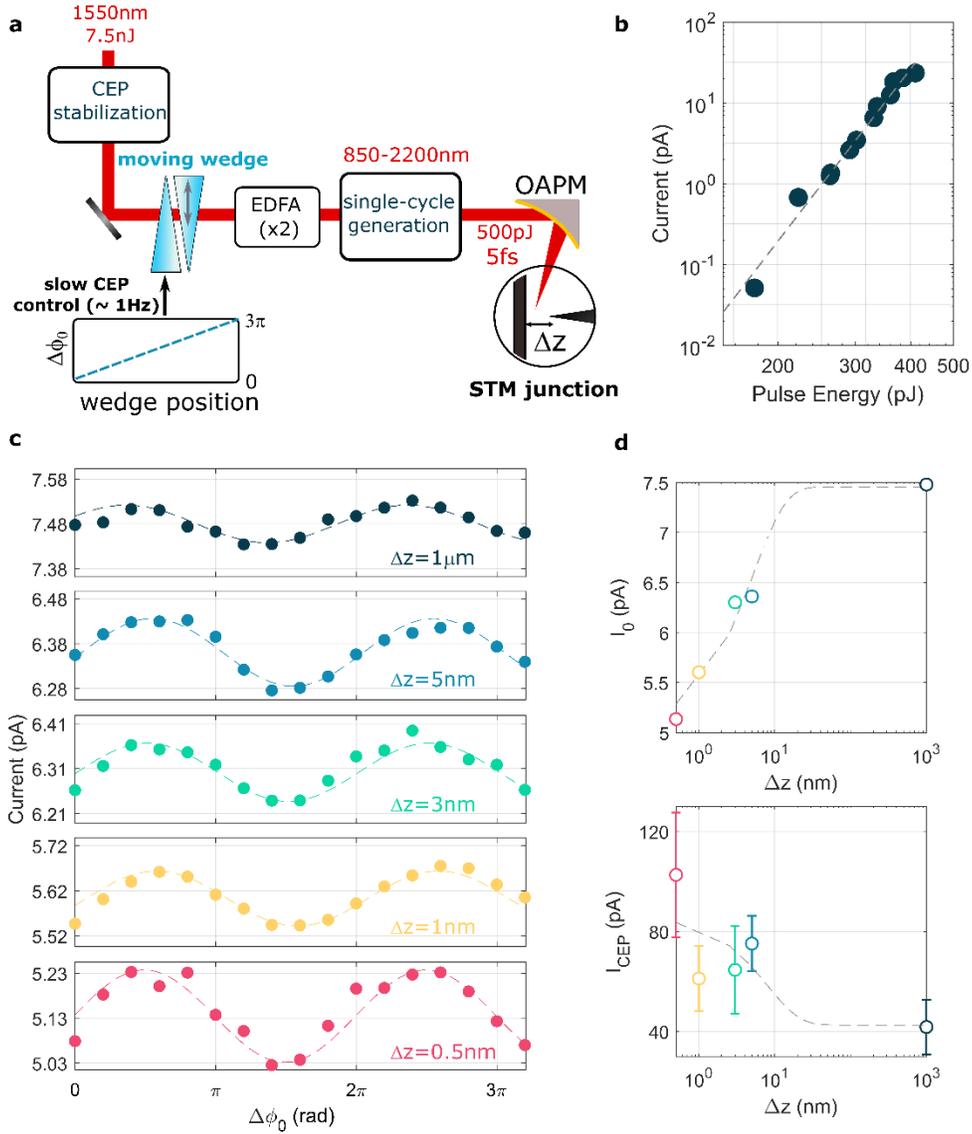

Figure 2: Gap-size dependence of coherent photoemission a) Sketch of the setup. b) Photocurrent from the isolated tip ($\Delta z = 1\ \mu m$) as function of the incoming pulse energy. The resulting graph is well reproduced by a 7-th order power law fit (dashed line), suggesting that a multi-photon process is the dominant photoemission mechanism. c) Photocurrent flowing across the STM junction for different values of the carrier-envelope phase (CEP) at different tip-sample distances. For each value of $\Delta z$, we fit the photocurrent with a sine function. d) Current offset $I_0$ (top) and CEP-modulated current amplitude $I_{CEP}$ (bottom) extracted from the best fit curves of panel c. An exponential fit (gray dashed line) serves for both cases as guide to the eye.

Starting from this condition, we disengage the STM feedback loop and set $V_0 = 0$ V, retract the sample from the tip by a known value $\Delta z$ (so that $d_{gap} = z_0 + \Delta z$) and start acquiring the photocurrent while the the CEP is scanned moving the insertion of the fused silica wedge. We repeat this procedure for different values of $\Delta z = 1\ \mu m, 5\ nm, 3\ nm, 1\ nm$ and $0.5\ nm$. The results of the measurements are shown in figure 2b. For all values of $d_{gap}$, we observe a sinusoidal modulation of the current with a $2\pi$ periodicity in the carrier-envelope phase (CEP), indicating that a portion of the STM current remains fully coherent with the electric field of the driving laser pulse. Remarkably, as the tip–sample distance is reduced from the isolated-tip regime ($\Delta z = 1\,\mu m$) to tunneling proximity ($\Delta z = 0.5\,nm$), the amplitude of the CEP-dependent current component doubles (bottom panel figure 2d), while the average total current decreases by approximately 30 % (top panel of figure 2d). Pinpointing the exact mechanisms governing the gap-dependence of the photocurrent is not trivial. Here we can reasonably argue that ponderomotive effects should be quenched in the tunnelling regime. We therefore attribute the enhancement of the CEP-dependent oscillation to the onset of field-driven tunnelling across the gap. This interpretation also explains the reduction of average current with decreasing gap size. At large gap sizes, photoemission is only possible for a specific field polarity (unidirectional emission from the tip) [42]. On the other hand, half-cycles of both polarities can contribute at smaller gaps, leading to full inversion of the CEP-dependent oscillation for a perfectly symmetric junction[19]. Despite the observed twofold enhancement of CEP-modulated amplitude at small gaps, the modulation depth is only around 2%. In the following, we focus on a readout strategy capable of isolating this minute signal while minimizing the required time for the measurement.

## 3. Electronic modulation of the carrier-envelope-phase

We recognize the observation of field-driven signatures in the STM current as a critical step toward realizing lightwave-driven STM at near-infrared frequencies, demonstrating that sub-cycle optical control of the tunneling current is achievable in this spectral range. Towards actual experiments, the ability to isolate sub-cycle coherent photocurrents from potentially largely dominant background signals is crucial. We will address this aspect this in the remainder of this work.

We propose to modulate the CEP and demodulate the current signal via lock-in detection at the CEP modulation frequency. Unlike intensity-based modulation techniques that rely on a chopper wheel, this approach fully rejects signals caused by spurious modulations of the thermal load impinging on the junction, resulting in a modulation of $d_{gap}$. Alternative modulation strategies have also been proposed, based for example on modulating the beam polarization or the duration of the field envelope [43]. While these approaches can be useful to discriminate laser-induced effects from a DC background, they cause strong spurious modulations of field waveform at the tunnelling junction, possibly making the detection of coherent effects even more complicated. On the other hand, the approach we propose directly isolates the CEP-dependent, i.e. coherent component of the photocurrent, providing a clean readout strategy for lightwave-driven STM experiments.
To achieve a lock-in detection bandwidth on the order of at least 1 Hz, the CEP must be modulated at frequencies exceeding 100 Hz ($\nu_{CEP} > 100\,Hz$). This requirement is incompatible with mechanical means of CEP control such as the wedge translation we used above. Instead, we propose to modulate the pump current of our preamplifier to achieve high-speed CEP modulation. In detail, we modulate the pump current of the preamplifier (EDFA1 in figure 3a) after the passive CEP stabilization. The modulation of the laser diode (LD) output power $I_{mod}(t)$ is imprinted onto the effective refractive index experienced by the laser pulses during amplification, ultimately affecting the CEP at the laser output.

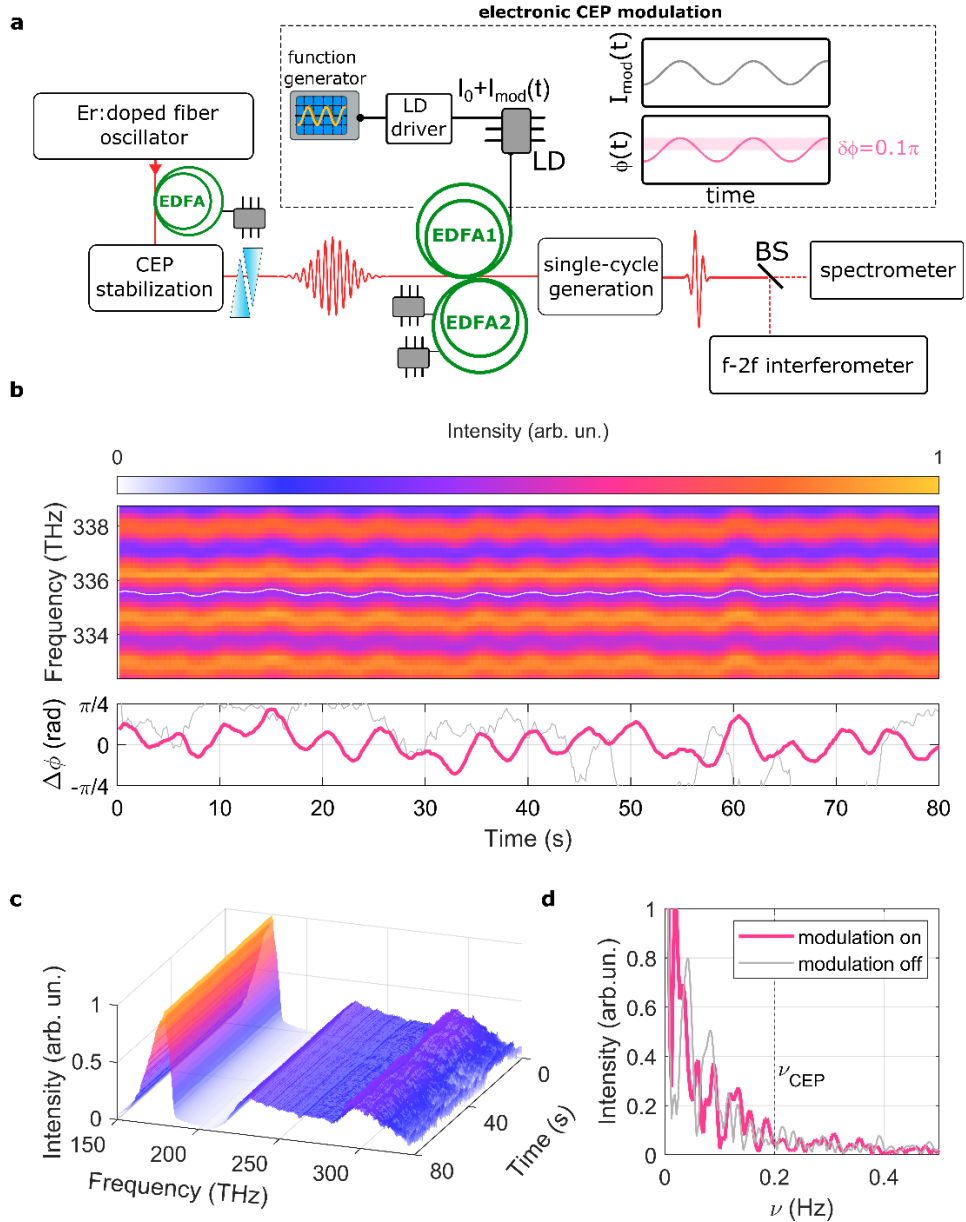

Figure 3: Electronic modulation of the CEP. a) Sketch of the setup. The CEP is controlled by modulating the pump current of the first Er:fiber amplifier (EDFA). b) (top panel) f-2f interferometer spectra acquired during the CEP modulation. The white line shows the extracted CEP shift profile. The graph is scaled to highlight the relation between the fringe modulation in the f-2f spectra and the extracted CEP shift. (bottom) CEP shift $\Delta\phi$ extracted from the f-2f spectra (pink curve). For comparison the measurement was repeated with the electronic CEP modulation switched off (gray curve). c) laser output spectrum recorded simultaneously with the data of panel b. e) Fourier analysis of the integrated spectra of panel c (pink curve). For comparison the same measurement was repeated with the electronic CEP modulation switched off (gray curve). The dashed vertical line indicates the CEP modulation frequency $\nu_{CEP}$. BS: beam splitter, LD: laser diode.

While a modulation of a pump current inherently also modulates the laser output power after EDFA1, power modulations of the single cycle synthesis should be largely suppressed by driving the second amplification stage (EDFA2 in figure 3a) in saturation. In the following, we will refer to this approach as the 'electronic CEP modulation'. It offers the possibility for long-term measurements since it is compatible with keeping the STM feedback loop engaged: While the wedge-based mechanical CEP modulation is limited to modulation frequencies of only few Hz, the feedback controlling $d_{gap}$ tries to maintain a constant current at a bandwidth $\simeq 100$ Hz. Therefore, slow modulations of the current would be effectively eliminated by the loop through a periodic re-adjustment of the gap size. Instead, a fast modulation of the CEP allows to modulate the currents faster than the feedback loop could compensate for.

In the following we first characterize the proposed electronic CEP modulation scheme, as sketched in figure 3a. While later measurements at the STM junction will be performed at $\nu_{CEP} > 100$ Hz, here we set $\nu_{CEP} = 0.2$ Hz to precisely track the phase modulation using an f-2f interferometer (top panel in figure 3b). As shown in the supporting information (figure S1), we observe that the change in CEP $\Delta\phi$ linearly follows $I_{mod}(t)$ and that an 8 % modulation of the pump current signal yields a $2\pi$ phase shift. To minimize the risk of spurious modulations of the pulse train at the laser output, we choose to modulate the CEP with a $0.2\pi$ peak-to-peak sinusoidal profile. In the bottom panel of figure 3b, we plot the phase shift $\Delta\phi$ extracted from the f-2f spectra. This phase modulation amplitude is comparable with the intrinsic CEP noise of our source in absence of an active stabilization loop (gray curve in the bottom panel of figure 3b). Simultaneously to the f-2f measurement, we monitor the laser output spectrum using a CCD spectrometer (figure 3c). Despite the high nonlinearity (third order) of the supercontinuum generation process, we do not observe quantifiable effects of the current-based CEP modulation on the supercontinuum spectrum. To exclude the presence of spurious modulations of the output power, in figure 3d we depict the integrated spectral intensity in the Fourier domain. For reference, we compare this with the intrinsic power noise of the system, obtained when the electronic CEP modulation is switched off. The two graphs do not show any appreciable difference, further demonstrating that the electronic CEP modulation does not introduce spurious modulations of the single-cycle waveform that are quantifiable compared the intrinsic noise of the laser system
.

## 4. Direct isolation of coherent photocurrents

After having characterized the electronic CEP modulation, we apply it to detect coherent photocurrents across the STM junction. The linear relation between the diode pump current and the CEP shift (figure SI1) opens several possibilities on how to implement the CEP modulation at the STM junction. Here we propose to isolate sub-cycle coherent photoemission by experimentally sampling the derivative of the photocurrent with respect to the CEP. We achieve this exploiting both the wedge and electronic CEP control, as schematically shown in figure 4a. Analogously to the measurements discussed in figure 1c, we slowly sweep the position of one of the wedges while acquiring the photocurrent. In the following, we will call $\phi_0$ the variation of CEP due to the wedge movement. We simultaneously introduce an electronic sinusoidal CEP modulation $\phi(t) = \delta\phi \sin(2\pi\nu_{CEP} t)$ with frequency $\nu_{CEP} > 100$ Hz and amplitude $\delta\phi =$

$0.1\pi$. Since the wedge scan is much slower than the electronic modulation, the instantaneous value of CEP can be written as $\text{CEP}(t) = \phi_0 + \phi(t)$.

Taking advantage of the small modulation amplitude $\delta\phi$ and neglecting a constant offset, a Taylor expansion of the photocurrent results in

$$I(\phi_0 + \delta\phi \sin(2\pi\nu_{CEP}t)) \propto \left.\frac{dI}{d\phi}\right|_{CEP=\phi_0} \sin(2\pi\nu_{CEP}t) \quad (1).$$

Within a first-order of approximation, the photocurrent follows a sinusoidal oscillation at the CEP modulation frequency, with a modulation amplitude that is proportional to the instantaneous value of its phase derivative. A pictorial representation of this "derivative sampling" is shown in figure 4b, where the cases $\phi_0 = (2n+1)\pi/2$ for n integer (maximum derivative signal) and $\phi_0 = n\pi$ ($dI/d\phi \simeq 0$) are highlighted. We feed the current signal into a lock-in amplifier and demodulate the signal at the CEP modulation frequency.

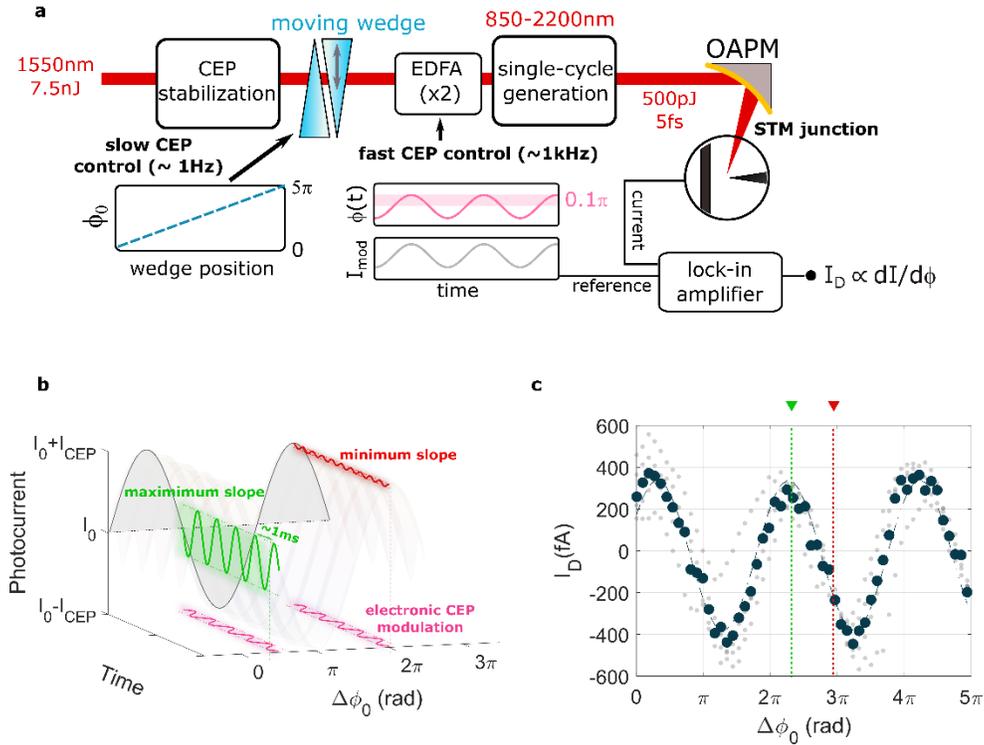

**Figure 4:** Isolation of phase-dependent photocurrents using the derivative sampling method. a) Schematic representation of the experiment. b) Pictorial representation of the derivative sampling method. The CEP is modulated using a $0.1\pi$ sinusoidal phase profile around $\phi_0$ (pink oscillating line). The value of $\phi_0$ is varied moving a pair of wedges. The modulated photocurrent (green and red curves) oscillates with an amplitude proportional to its derivative at the value $\phi_0$ set by the wedge insertion. c) Experimental demodulated photocurrent (dark blue circles) from the isolated tip, averaged over four successive CEP sweeps. The light gray circles show the current from the individual scans. The averaged curve is fitted with a sine function (dashed line), shown as guide to the eye. The green and red vertical lines mark the conditions for maximum and minimum modulation of the photocurrent.

As shown by equation 1, acquiring the lock-in output $I_D$ at different values of wedge insertion we effectively sample the derivative of the photocurrent with respect to the CEP. In this way, we expect to capture only coherent sub-cycle contributions, since all phase-independent photocurrents are inherently rejected. To validate this approach, in figure 4c we plot the demodulated photocurrent from the isolated STM tip as the value of $\phi_0$ is scanned over roughly $5\pi$. The photocurrent is averaged over four successive scans, corresponding to a one-order-of-magnitude reduction in acquisition time compared to the data discussed in figure 2b. The demodulated photocurrent shows a clean oscillation coherent with the CEP shift set by the wedges, analogously to the data discussed in figure 2b. We attribute the difference in oscillation amplitudes between the dataset of figure 2 and figure 4 to a combination of different readout mechanism and slightly different beam-tip coupling conditions, due to the extreme nonlinearity of the process. Independently, while the total photocurrent only featured a small modulation amplitude around 0.5 % in the retracted case (figure 2b, top panel), the demodulated signal remarkably shows 100 % modulation with full inversion around zero. This experimentally demonstrates that coherent sub-cycle photocurrents can be completely isolated from a dominating incoherent background using the derivative sampling method.

## 5. Conclusions

In this work, we investigated coherent photoemission across the junction of a scanning tunnelling microscope illuminated by CEP-stable single-cycle near-IR laser transients. To quantify the magnitude of field-driven effects at the junction, we first characterized its CEP dependence based on a slow mechanical modulation technique that is known to cause no spurious artifacts. We observed clear modulation of the photocurrent consistent with the CEP sweep, demonstrating the possibility of coherent control of photoemission across the STM junction. We then investigated the effect of gap size on coherent photoemission and observed an increase of the CEP-dependent oscillations when the gap size is decreased down to the tunnelling regime. These findings show that a lightwave STM at near-IR is viable. However, coherent photocurrents represent only a small fraction of the total signal, in practice rendering the detection of sub-cycle dynamics in an actual experiment extremely challenging. To address this fact, we proposed to isolate coherent photocurrents by lock-in detection based on a fast electronic modulation of the CEP at frequencies $\nu_{CEP} > 100$ Hz. After characterizing this configuration, we combined it with the slow mechanical CEP control to define the operating points of the lock-in detection. By integrating a slow mechanical CEP sweep with a fast and 'perturbative' electronic modulation, we demonstrate the possibility to capture the coherent signal directly by sampling its derivative. Considering our previous observation of a 0.5 % modulation amplitude for the total photocurrent, the derivative sampling yields a 200-fold enhancement of modulation depth (and therefore of sensitivity in isolating coherent effects), together with a one-order-of magnitude reduction in acquisition time. Moreover, the derivative sampling method fully rejects intensity-dependent spurious signals from the detected photocurrent. Finally, it can be performed with the STM feedback loop engaged, as long as sufficiently fast CEP modulation frequencies are chosen. This makes the detection of sub-cycle dynamics compatible with the high stability required at the STM junction. We therefore believe that the readout scheme proposed in this work provides a robust and practical solution to implement future near-IR-driven ultrafast STM experiments with sub-optical-cycle time resolution.

**Research funding:** D.B., A.R., F.P, M.A and M.L. acknowledge funding by the European Research Council (819871) and the FEDER Program (Grant No. 2017-03-022-19 "Lux-Ultra-Fast" and Grant No. 2023-01-04 "Lux-Ultra-Fast 2"). ML acknowledges funding from the Institute for Advanced Studies of the University of Luxembourg, and the European Union's Horizon 2020 research and innovation programme under the Marie Skłodowska-Curie Actions (GA #101081455). A.L. acknowledges funding by Deutsche Forschungsgemeinschaft, Project-ID 425217212 (SFB 1432).

**Acknowledgements:** The authors thank Melanie Müller for insightful discussions that improved the quality of this manuscript.

**Disclosures**: The authors declare no conflicts of interest.

**Data availability:** The datasets analyzed during the current study are available from the corresponding authors upon reasonable request

†**current affiliation:** Department of Physical Chemistry, Fritz Haber Institute of the Max Planck Society, 14195 Berlin, Germany

**References:**

[1]  A. Grupp *et al.*, "Broadly tunable ultrafast pump-probe system operating at multi-kHz repetition rate," *Journal of Optics (United Kingdom)*, vol. 20, no. 1, Jan. 2018, doi: 10.1088/2040-8986/aa9b07.

[2]  M. Ortolani, A. Mancini, A. Budweg, D. Garoli, D. Brida, and F. De Angelis, "Pump-probe spectroscopy study of ultrafast temperature dynamics in nanoporous gold," *Phys Rev B*, vol. 99, no. 3, Jan. 2019, doi: 10.1103/PhysRevB.99.035435.

[3]  C. O. Karaman, A. Y. Bykov, F. Kiani, G. Tagliabue, and A. V. Zayats, "Ultrafast hot-carrier dynamics in ultrathin monocrystalline gold," *Nat Commun*, vol. 15, no. 1, Dec. 2024, doi: 10.1038/s41467-024-44769-3.

[4]  J. Allerbeck, T. Deckert, L. Spitzner, and D. Brida, "Probing free-carrier and exciton dynamics in a bulk semiconductor with two-dimensional electronic spectroscopy," *Phys Rev B*, vol. 104, no. 20, Nov. 2021, doi: 10.1103/PhysRevB.104.L201202.

[5]  E. Beaurepaire, J.-C. Merle, A. Daunois, and J.-Y. Bigot, "Ultrafast Spin Dynamics in Ferromagnetic Nickel," 1996.

[6]  T. L. Cocker, V. Jelic, R. Hillenbrand, and F. A. Hegmann, "Nanoscale terahertz scanning probe microscopy," Aug. 01, 2021, *Nature Research*. doi: 10.1038/s41566-021-00835-6.

[7] T. Siday *et al.*, "All-optical subcycle microscopy on atomic length scales," *Nature*, vol. 629, no. 8011, pp. 329–334, May 2024, doi: 10.1038/s41586-024-07355-7.

[8] M. Wagner *et al.*, "Ultrafast dynamics of surface plasmons in InAs by time-resolved infrared nanospectroscopy," *Nano Lett*, vol. 14, no. 8, pp. 4529–4534, Aug. 2014, doi: 10.1021/nl501558t.

[9] M. Wagner *et al.*, "Ultrafast and nanoscale plasmonic phenomena in exfoliated graphene revealed by infrared pump-probe nanoscopy," *Nano Lett*, vol. 14, no. 2, pp. 894–900, Feb. 2014, doi: 10.1021/nl4042577.

[10] M. Müller, "Imaging surfaces at the space–time limit: New perspectives of time-resolved scanning tunneling microscopy for ultrafast surface science," *Prog Surf Sci*, no. xxxx, p. 100727, 2023, doi: 10.1016/j.progsurf.2023.100727.

[11] A. G. Borisov, B. Ma, M. Zapata-Herrera, A. Babaze, M. Krüger, and J. Aizpurua, "Femtosecond Optical-Field-Driven Currents in Few-Nanometer-Size Gaps with Hot Electron Injection into Metallic Leads," *ACS Photonics*, 2025, doi: 10.1021/acsphotonics.4c02612.

[12] T. L. Cocker *et al.*, "An ultrafast terahertz scanning tunnelling microscope," *Nat Photonics*, vol. 7, no. 8, pp. 620–625, 2013, doi: 10.1038/nphoton.2013.151.

[13] T. L. Cocker, D. Peller, P. Yu, J. Repp, and R. Huber, "Tracking the ultrafast motion of a single molecule by femtosecond orbital imaging," *Nature*, vol. 539, no. 7628, pp. 263–267, 2016, doi: 10.1038/nature19816.

[14] C. Andrea Rozzi *et al.*, "Quantum coherence controls the charge separation in a prototypical artificial light-harvesting system," *Nat Commun*, vol. 4, 2013, doi: 10.1038/ncomms2603.

[15] Y. Arashida *et al.*, "Subcycle Mid-Infrared Electric-Field-Driven Scanning Tunneling Microscopy with a Time Resolution Higher Than 30 fs," 2022, doi: 10.1021/acsphotonics.2c00995.

[16] M. Garg and K. Kern, "Attosecond coherent manipulation of electrons in tunneling microscopy," *Science (1979)*, vol. 1098, no. November, p. eaaz1098, 2019, doi: 10.1126/science.aaz1098.

[17]   G. Krauss et al., "Synthesis of a single cycle of light with compact erbium-doped fibre technology," *Nat Photonics*, vol. 4, no. 1, pp. 33–36, Jan. 2010, doi: 10.1038/nphoton.2009.258.

[18]   M. Ludwig et al., "Sub-femtosecond electron transport in a nanoscale gap," *Nat Phys*, vol. 16, no. 3, pp. 341–345, 2020, doi: 10.1038/s41567-019-0745-8.

[19]   T. Rybka, M. Ludwig, M. F. Schmalz, V. Knittel, D. Brida, and A. Leitenstorfer, "Sub-cycle optical phase control of nanotunnelling in the single-electron regime," *Nat Photonics*, vol. 10, no. 10, pp. 667–670, 2016, doi: 10.1038/nphoton.2016.174.

[20]   T. L. Cocker et al., "An ultrafast terahertz scanning tunnelling microscope," *Nat Photonics*, vol. 7, no. 8, pp. 620–625, 2013, doi: 10.1038/nphoton.2013.151.

[21]   K. Yoshioka et al., "Real-space coherent manipulation of electrons in a single tunnel junction by single-cycle terahertz electric fields," Nov. 29, 2016, *Nature Publishing Group*. doi: 10.1038/nphoton.2016.205.

[22]   C. Lin et al., "Continuous-Wave Multiphoton-Induced Electron Transfer in Tunnel Junctions Driven by Intense Plasmonic Fields," *ACS Photonics*, vol. 10, no. 10, pp. 3637–3646, 2023, doi: 10.1021/acsphotonics.3c00714.

[23]   B. Schröder et al., "Controlling photocurrent channels in scanning tunneling microscopy," *New J Phys*, vol. 22, no. 3, 2020, doi: 10.1088/1367-2630/ab74ac.

[24]   S. Liu, A. Hammud, I. Hamada, M. Wolf, M. Müller, and T. Kumagai, "Nanoscale coherent phonon spectroscopy," 2022. [Online]. Available: https://www.science.org

[25]   N. Martín Sabanés, F. Krecinic, T. Kumagai, F. Schulz, M. Wolf, and M. Müller, "Femtosecond Thermal and Nonthermal Hot Electron Tunneling Inside a Photoexcited Tunnel Junction," *ACS Nano*, vol. 16, no. 9, pp. 14479–14489, 2022, doi: 10.1021/acsnano.2c04846.

[26]   S. Thomas, M. Krüger, M. Förster, M. Schenk, and P. Hommelhoff, "Probing of optical near-fields by electron rescattering on the 1 nm scale," *Nano Lett*, vol. 13, no. 10, pp. 4790–4794, 2013, doi: 10.1021/nl402407r.

[27]     S. Grafström, "Photoassisted scanning tunneling microscopy," *J Appl Phys*, vol. 91, no. 3, pp. 1717–1753, 2002, doi: 10.1063/1.1432113.

[28]     V. Gerstner, A. Thon, and W. Pfeiffer, "Thermal effects in pulsed laser assisted scanning tunneling microscopy," *J Appl Phys*, vol. 87, no. 5, pp. 2574–2580, 2000, doi: 10.1063/1.372221.

[29]     S. Grafström, P. Schuller, J. Kowalski, and R. Neumann, "Thermal expansion of scanning tunneling microscopy tips under laser illumination," *J Appl Phys*, vol. 83, no. 7, pp. 3453–3460, 1998, doi: 10.1063/1.366556.

[30]     R. Huber, M. Koch, and J. Feldmann, "Laser-induced thermal expansion of a scanning tunneling microscope tip measured with an atomic force microscope cantilever," *Appl Phys Lett*, vol. 73, no. 17, pp. 2521–2523, 1998, doi: 10.1063/1.122502.

[31]     B. Piglosiewicz *et al.*, "Carrier-envelope phase effects on the strong-field photoemission of electrons from metallic nanostructures," *Nat Photonics*, vol. 8, no. 1, pp. 37–42, 2014, doi: 10.1038/nphoton.2013.288.

[32]     M. Krüger, M. Schenk, M. Förster, and P. Hommelhoff, "Attosecond physics in photoemission from a metal nanotip," *Journal of Physics B: Atomic, Molecular and Optical Physics*, vol. 45, no. 7, 2012, doi: 10.1088/0953-4075/45/7/074006.

[33]     D. Brida, G. Krauss, A. Sell, and A. Leitenstorfer, "Ultrabroadband Er: Fiber lasers," *Laser Photon Rev*, vol. 8, no. 3, pp. 409–428, 2014, doi: 10.1002/lpor.201300194.

[34]     C. Schoenfeld, P. Sulzer, D. Brida, A. Leitenstorfer, and T. Kurihara, "Passively phase-locked Er:fiber source of single-cycle pulses in the near infrared with electro-optic timing modulation for field-resolved electron control," *Opt Lett*, vol. 47, no. 14, p. 3552, Jul. 2022, doi: 10.1364/ol.461076.

[35]     J. R. Birge, R. Ell, and F. X. Kärtner, "Two-dimensional spectral shearing interferometry for few-cycle pulse characterization and optimization," *Springer Series in Chemical Physics*, vol. 88, pp. 160–162, 2007, doi: 10.1007/978-3-540-68781-8_51.

[36]     A. Rossetti, M. Falk, A. Leitenstorfer, D. Brida, and M. Ludwig, "Gouy phase effects on photocurrents in plasmonic nanogaps driven by single-cycle

pulses," *Nanophotonics*, vol. 13, no. 15, pp. 2803–2809, Jul. 2024, doi: 10.1515/nanoph-2023-0897.

[37] N. Behr and M. B. Raschke, "Optical antenna properties of scanning probe tips: Plasmonic light scattering, tip-sample coupling, and near-field enhancement," *Journal of Physical Chemistry C*, vol. 112, no. 10, pp. 3766–3773, 2008, doi: 10.1021/jp7098009.

[38] M. Y. Ivanov, M. Spanner, and O. Smirnova, "Anatomy of strong field ionization," *J Mod Opt*, vol. 52, no. 2–3, pp. 165–184, 2005, doi: 10.1080/0950034042000275360.

[39] M. Kaack and D. Fick, "Determination of the work functions of Pt(lll) and Ir(lll) beyond 1100 K surface temperature," 1995.

[40] C. Lin *et al.*, "Continuous-Wave Multiphoton-Induced Electron Transfer in Tunnel Junctions Driven by Intense Plasmonic Fields," *ACS Photonics*, vol. 10, no. 10, pp. 3637–3646, 2023, doi: 10.1021/acsphotonics.3c00714.

[41] G. Herink, D. R. Solli, M. Gulde, and C. Ropers, "Field-driven photoemission from nanostructures quenches the quiver motion," *Nature*, vol. 483, no. 7388, pp. 190–193, 2012, doi: 10.1038/nature10878.

[42] P. D. Keathley *et al.*, "Vanishing carrier-envelope-phase-sensitive response in optical-field photoemission from plasmonic nanoantennas," *Nat Phys*, vol. 15, no. 11, pp. 1128–1133, 2019, doi: 10.1038/s41567-019-0613-6.

[43] M. Garg, A. Martin-Jimenez, Y. Luo, and K. Kern, "Ultrafast Photon-Induced Tunneling Microscopy," *ACS Nano*, vol. 15, no. 11, pp. 18071–18084, 2021, doi: 10.1021/acsnano.1c06716.

# Supporting information

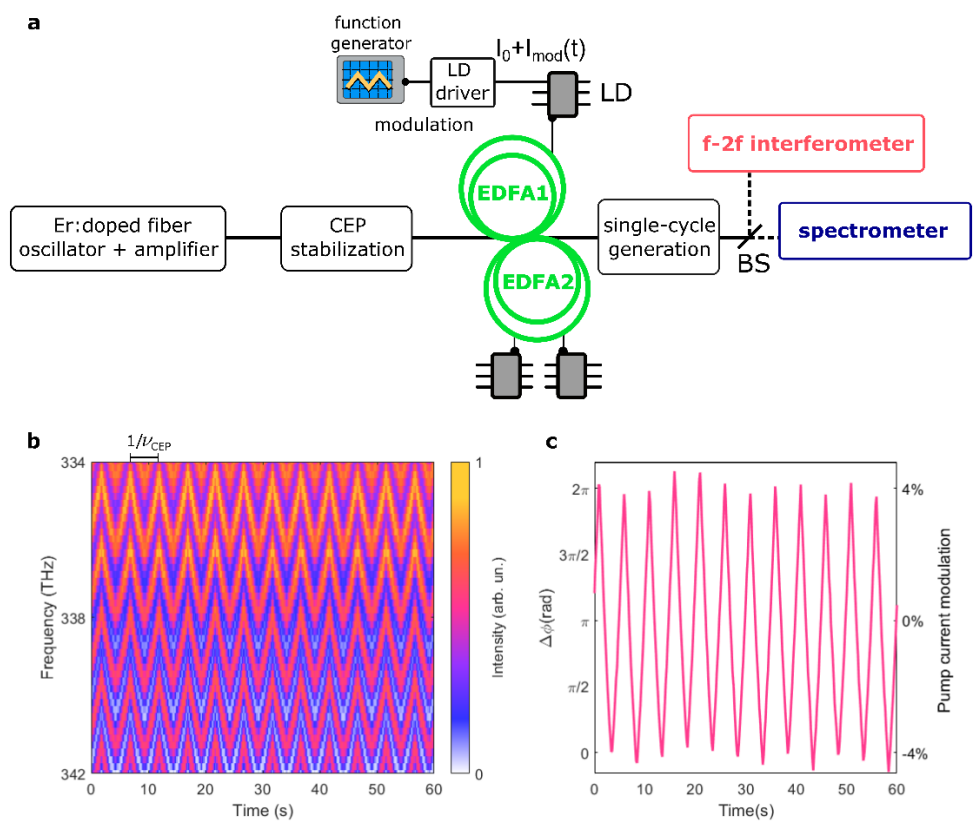

Figure S1: Characterization of the CEP shift induced by modulating the pump current of the laser diode (LD) pumping the preamplifier (EDFA1). a) f-2f spectra acquired during a 8% triangular modulation of the LD pump currents. c) extracted phase shift, showing the a 2π peak-to-peak modulation amplitude.